# Measurement of ultrashort optical pulses via time lens imaging in CMOS compatible waveguides

Alessia Pasquazi, *Member IEEE*, Yongwoo Park, Sai T.Chu, Brent .E. Little, François Légaré, Roberto Morandotti, *Member IEEE*, José Azaña, *Member IEEE,* and David J. Moss, *Senior Member IEEE*

*Abstract*— We demonstrate temporal measurements of subpicosecond optical pulses via time-to-frequency conversion in a 45cm long CMOS compatible high index glass spiral waveguide. The measurements are based on efficient four wave mixing in the C-band, using around 1W of peak pump power. We achieve a resolution of 400fs over a time window of 100ps, representing a time-bandwidth product > 250.

I. INTRODUCTION

Signal measurement and monitoring play a key role in the development of fiber optics telecommunication systems. The push towards data stream speeds of the order of terabits per seconds [1]-[3] requires real-time methods for imaging optical pulses with subpicosecond resolution. However electronic-based measurement tools (e.g. photodetection followed by an electronic oscilloscope) are only suitable for characterizing optical signal longer than tens of picoseconds. All-optical approaches based on ultrafast optical nonlinearities have been successfully employed for optical signal processing. When implemented in a waveguide geometry, they provide useful instruments able to overcome the bandwidth limitations associated with electronic approaches [4]-[12].

In particular, ultrafast nonlinearities can be conveniently exploited for temporal signal imaging through the so-called space-time duality for ultrafast optical processing [13]-[22]. In simple terms, refractive optical elements such as lenses and prisms have temporal analogs that perform equivalent functions in the time domain, as they are represented by the same mathematical equations.

Manuscript received November 28, 2010; revised February 2, 2011; accepted February 14, 2011. Date of publication April 7, 2011; date of current version March 2, 2012. This work was supported in part by the National Research Council of Canada, in part by the Fonds Québécois de la Recherche sur la Nature et les Technologies, and in part by the Australian Research Council.

A. Pasquazi, Y. Park, F. Légaré, R. Morandotti, and J. Azaña are with the Institut National de la Recherche Scientifique—Centre Energie, Matériaux et Télécommunications (INRS-EMT), Université du Québec, Montreal, QC J3X 1S2 Canada.
B. E. Little is with Infinera Corporation, Annapolis Jct, MD 94089 USA.
S. T. Chu is with the Department of Physics and Material Science, City University of Hong Kong, Kowloon, Hong Kong.
D. J. Moss is with the Center for Ultra-high Bandwidth Devices for Optical Systems (CUDOS), School of Physics, University of Sydney, New SouthWales 2006, Australia, and also with INRS-EMT, Université du Québec, Montreal, Québec J3X 1S2 Canada (dmoss@physics.usyd.edu.au).

Just as a spatial imaging system can magnify or reduce an image, a "temporal imaging system" can temporally broaden or compress an optical waveform in time. Similarly, as a spatial lens can be used to obtain the Fourier transform (FT) of a spatial profile, a "time lens" (TL) can be used to obtain the FT of a temporal profile, consequently implementing a time to frequency conversion operation [19]-[20]. More specifically, we define a TL as a physical mechanism that induces a quadratic phase profile on an input temporal optical signal, in analogy to a spatial lens, which adds a quadratic curvature to the phase front of a transmitted radiation. The importance of this equivalence is associated with the fact that a vast category of optical instruments for diffractive beams are based on a suitable combination of lenses and free-space beam propagation. Thus, equivalent concepts can be employed for signal processing temporal signals by properly combining TLs and dispersive elements, since the Fresnel equation [15] models both pulse propagation in a group-velocity dispersive system and free-space beam propagation.

In the optical domain, the TL concept was first implemented using an electro-optic phase modulator driven by a sinusoidal RF signal [15], [16], [20].

This solution is however limited to processing optical waveforms with time features longer than ~10ps. Later on, it was suggested that this limitation could be overcome by resorting to suitable ultrafast nonlinear processes.

In particular, the TL bandwidth was increased by using cross-phase modulation (XPM) with parabolic pulses to generate the required quadratic phase shift [23]. However, in this configuration the equivalent focal length was limited in practice by the pump power. The best results to date for a TL process have been obtained by using a parametric process, such as three wave mixing or degenerate four wave mixing (TWM-FWM) [17], [18], [21], [24]. Here the processed waveform is obtained from the output idler after mixing the signal of the interaction with a conveniently dispersed Gaussian pump, as depicted in Fig. 1. In contrast to the XPM configuration, in the FWM based approach the focal length is determined by the amount of dispersion of the pump (i.e. the chirp rate of the pump). As such [25] the pump power only influences the conversion efficiency of the signal into the idler, without affecting the focal length of the TL.

The potential of this new TL approach for ultrashort pulse characterization [21] was first demonstrated by using a nonlinear TWM process in bulk quadratic crystals. More recently an integrated implementation of a TL was proposed in silicon nanowires via degenerate FWM [18]. This approach has the advantage of being suitable for centro-symmetric materials, such as silicon or glass – both being fundamental platforms for integrated optics. In general, the FWM based approach tends to be sensitive to deviations from the optimal nonlinear conditions, and in particular can be limited by nonlinear absorption at higher pump powers. For this reason there is a strong motivation to explore novel material platforms that exhibit high nonlinearity with low linear and nonlinear losses in order to optimize the all-optical interactions.

Platforms compatible with electronic integration technology (CMOS) are desirable in this scenario, both to reduce the manufacturing costs as well as to potentially integrate electronic and photonic functionalities on the same chip.

In this paper, we exploit high index doped silica glass as a platform to demonstrate time to frequency conversion via FWM TL in a spiral waveguide. Our waveguides are quite attractive for Kerr based operation in the C-Band [25]-[29]. FWM in our waveguides shows levels of gain and bandwidth comparable to the best results in chalcogenide waveguides [12], [30]. In contrast, both chalcogenides and silicon nanowires, can be affected by nonlinear absorption, whereas high index doped silica glass shows no intensity dependent saturation [7], [30], [31]. This platform can also be efficiently interfaced with single mode fibers (SMFs), since it possesses a relatively low index mismatch with SMFs, and it does not require the nanowire dimensions that silicon, for example, needs in order to achieve efficient FWM [30]. Indeed, by incorporating on-chip mode field converters, typical pigtail coupling losses to SMF are less than 1.5dB.

Based on this capability, here we demonstrate time-to frequency conversion of subpicosecond optical pulses over a 100ps time window, obtaining an instrument for imaging the time-domain intensity profile of ultrafast pulses with a remarkable time bandwidth product of 250.

## II. HIGH INDEX GLASS WAVEGUIDES

The device under investigation is a 45cm long spiral waveguide with a rectangular cross section core of 1.45 μm x 1.50 μm composed of high index doped silica glass [32], [33] ($n_{core}$= 1.7 @ 1550nm) surrounded by silica, on a silicon wafer (see Fig. 2). The layers were deposited by chemical vapor deposition and the spiral was patterned with high resolution optical lithography followed by reactive ion etching.

The 45cm long spiral waveguide (see Fig. 3) is contained within a square chip area as small as 2.5mm x 2.5mm and is pigtailed to SMFs. The properties of the material and waveguide dimensions were engineered to reduce the material dispersion near λ=1550nm, with an anomalous group velocity dispersion $\beta_2$ of the order of 10 ps$^2$/km in the middle of the C-band (see Fig. 4), thus enabling a large wavelength range for FWM phase matching. The nonlinearity parameter (γ) in these waveguides is 220W$^{-1}$km$^{-1}$ with the Kerr nonlinearity ($n_2$) being about 5x that of silica glass.

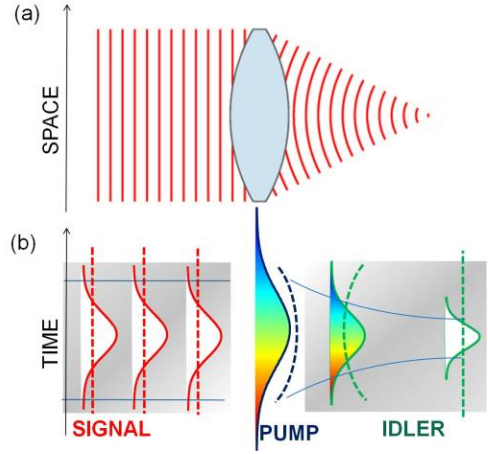

Fig. 1 Principle of a TL based on parametric wave mixing. (a) A collimated beam focused by a lens in the space domain. (b) The equivalent operation in the temporal domain performed by a wave mixing TL. The signal of the parametric interaction plays the role of the input beam to the lens. It interacts with a dispersed Gaussian pump generating a frequency converted idler. The quadratic chirp and the temporal duration of the pump define, respectively, the focal length and the numerical aperture of the TL. The generated idler acquires the signal profile and the pump chirp and behaves as the transmitted beam from a lens.

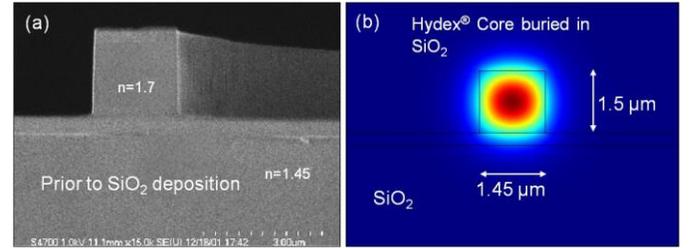

Fig. 2 Cross-section of high index glass waveguide. SEM image (a) and mode profile (b).

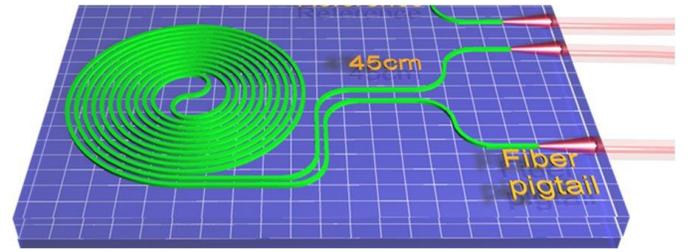

Fig. 3 Schematic of a 45cm long spiral waveguide structure.

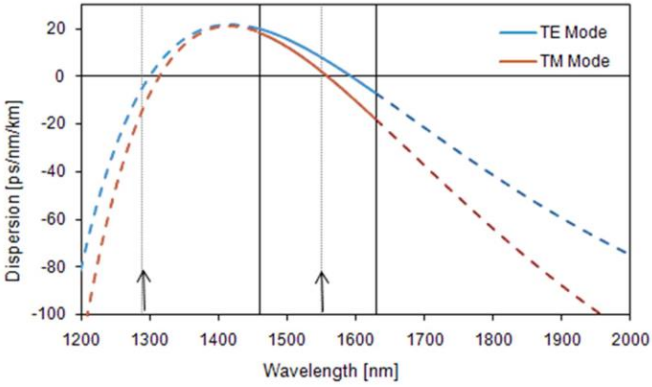

**Fig. 4** Dispersion of the doped silica glass waveguide for TE and TM polarizations. The solid curves between 1470nm and 1620nm represent experimental measurements, while the dashed curves are theoretical extrapolations. (after Ref.[31])

### III. FOUR WAVE MIXING PERFORMANCE

In order to demonstrate the suitability of our platform for integrated TL based operation, we experimentally characterized the FWM performance of the spiral waveguide [28]. Fig. 5 shows the results we obtained by using subpicosecond pulses for both the pump and the signal. The measurements were carried out with an OPO system that generates 180fs (bandwidth = 30nm) long pulses at a repetition rate of 80MHz. The broadband pulse source was split and filtered by way of two tunable Gaussian bandpass filters operating in transmission with a -3dB bandwidth of 5nm (equivalent to a pulsewidth of ~700fs), in order to obtain synchronized and coherent pump and signal pulses at two different wavelengths. The pump and signal pulses were combined into a standard SMF using a (90/10)% beam splitter and then coupled into the spiral waveguide. Pulse synchronization was adjusted by means of an optical delay line, while power and polarization were controlled with a polarizer and a λ/2 plate. Both pump and probe polarizations were aligned to the quasi-TE mode of the device.

Fig. 5 shows the output spectra for different peak pump powers, with a pump wavelength of 1550nm and a signal wavelength of 1480nm. We define the "on/off" conversion efficiency from signal $s(t)$ to idler $i(t)$ of the process as the ratio of the transmitted pulse *energy*, rather than peak power, of the idler to the transmitted signal without the pump [31].

$$\eta = \lim_{T \to \infty} \frac{\int_{-T/2}^{T/2} |i(t)|^2 \, dt}{\int_{-T/2}^{T/2} |s(t)|^2 \, dt} \qquad (1)$$

This allows us to account for the effect of the pulses shape and for the spectral broadening due to XPM, which lowers the spectral intensity. The net "on-chip" gain is then the "on/off" gain minus the propagation loss. The experimental on/off efficiency vs. pump peak power [28] is shown in Fig. 5 (b) along with the theoretical calculations for both a continuous wave (CW) and a pulsed pump. The modeling and experiment agree quite well for the pulsed case. The CW case represents the maximum achievable gain for a given pump peak power, as it maximizes relation (1), and it is insensitive to detrimental effects such as spectral broadening and temporal walk-off that can limit the efficiency in the pulsed case. For a 38W pump power, we measured a maximum on/off FWM conversion efficiency of +16.5dB from signal to idler. This translates into a net on-chip conversion efficiency of +13.7dB and a gain of +12.3dB, when the overall propagation loss of 2.7dB is included. Furthermore, the low dispersion also results in a remarkably large bandwidth of almost 200nm (signal to idler) [28]. We note that even at the highest pump powers used in these experiments we do not observe any sign of saturation and this is largely due to the negligible nonlinear absorption in this platform (up to intensities of 25GW/cm$^2$ [33]). This FWM performance is adequate enough to serve as the basis for TL pulse measurement, to which we turn next.

### IV. TIME LENS AND TIME-TO-FREQUENCY CONVERSION

Temporal-to-frequency domain conversion is an interesting application of a TL. This mechanism has proven particularly useful for measuring the time-domain intensity profiles of ultrafast optical waveforms. In the spatial case it is well known that the optical field transmitted by a lens in its focal plane is the FT of the spatial input field distribution in the lens focus. In the temporal case, it has been previously shown [20] that a system consisting of a first-order dispersive medium (inducing a predominantly linear chirp on the incoming optical signal) followed by a TL can be configured to map the input temporal shape into the output spectrum. This allows one to capture the time-domain intensity profile through a simple optical spectrum measurement. In particular, the linear chirp induced by the dispersive line on the pulse under test (PUT) must be exactly compensated for by the subsequent TL process.

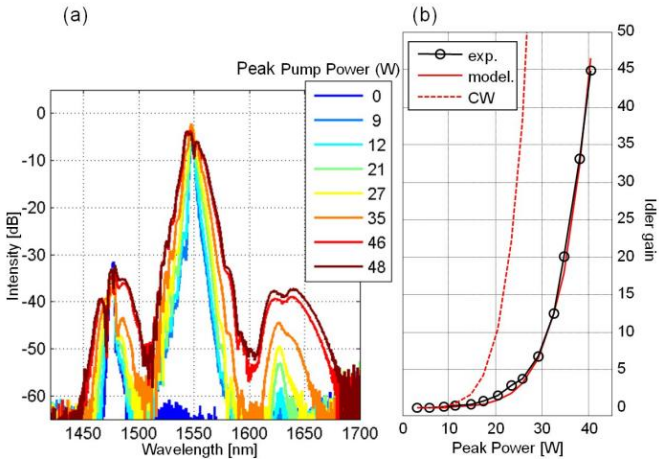

**Fig. 5** (a) Output spectrum for pulsed FWM measurements with a pump wavelength of 1550nm and a signal wavelength of 1480nm, for increasing peak pump powers (expressed in W). (b) Resulting on/off conversion efficiency as a function of the pump power, along with theoretical results for both pulsed and CW conditions (after Ref. [28])

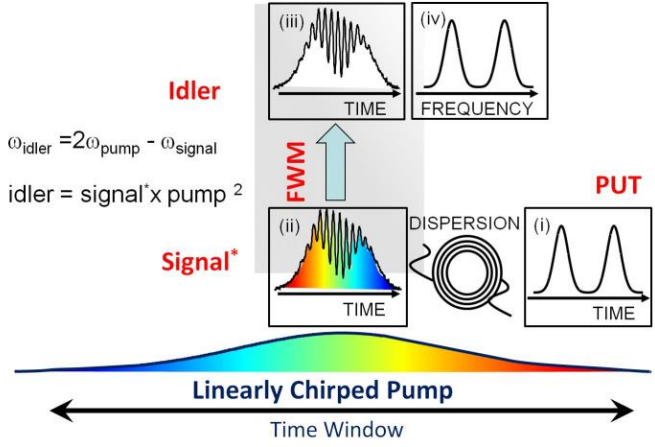

Fig. 6 – Principle of time-to-frequency conversion using TL based on the FWM experienced by the pulse under test (PUT) when interacting with a linearly chirped pump pulse. In the case illustrated here, the PUT first undergoes a sufficiently large dispersion to enter the so-called temporal Fraunhofer regime, where the time-domain optical waveform at the input of the TL (signal) is proportional to the spectral shape (amplitude variation) of the PUT and is phase chirped due to the large dispersion (i). When the signal interacts with the pump (ii) the nonlinear frequency conversion produces an idler with the temporal shape of the signal but with a chirp exactly compensated by the pump chirp (iii). The output of the TL (idler) is transform limited and its temporal shape is proportional to the spectrum of the PUT. As a result, the spectrum at the TL output is proportional to the temporal amplitude profile of the PUT (iv).

Practically speaking, a pulse acting as the pump for the parametric interaction is initially stretched with a dispersive element of length L and group velocity dispersion $\beta_2$ (i.e., a spool of SMF), resulting in a temporal phase curvature $\phi_P = \beta_2 L$. The TL effective phase curvature resulting from a FWM process is $\phi_F = \phi_P /2$, due to the quadratic dependence on the pump profile of the idler for a FWM interaction [18]. We note that in the case of TWM based TL, the condition would be $\phi_F = \phi_P$.[17],[21]

The PUT is dispersed by a total amount of predominantly first-order dispersion to match the TL phase curvature $\phi_F$, and is then coupled into the waveguide as the signal for the parametric interaction. This signal (i.e the PUT with the appropriate chirp applied) plays the role of the input waveform to the TL, and the idler behaves as the transmitted (output) waveform, since it acquires the signal temporal shape and the TL phase curvature $\phi_F$. In this configuration the idler energy spectrum maps the temporal intensity shape of the PUT according to the following time-to-frequency scaling law:

$$t = \phi_F \, \omega. \qquad (2)$$

A temporal imaging system is then realized by measuring the idler with a spectrometer. The temporal window of the TL based imaging system is equal to the temporal duration of the dispersed pump, while the resolution depends on the pump bandwidth. Assuming that $\tau_0$ is the duration for a Gaussian pump in its transform limited version, the resolution is approximately given by $\tau_0 / \sqrt{2}$ [18],[20].

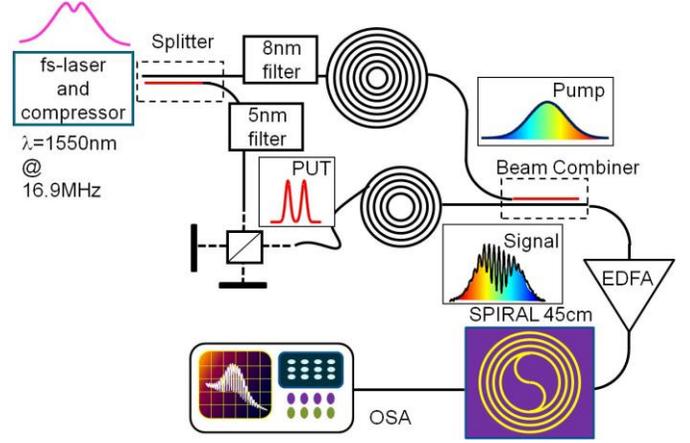

Fig. 7 – Schematic of time-lens measurement setup.

As an illustrative simple example (see Fig. 6), let us consider a transform-limited PUT. As previously mentioned, the 'signal' at the TL input is the temporally dispersed version of the PUT, and it follows in time the PUT spectral amplitude shape for a sufficiently large amount of dispersion (each frequency being associated with a different time coordinate, i.e. the when the temporal Fraunhofer condition is satisfied), and the phase is chirped. It is important to underline that the latter *is not a necessary condition* to achieve the desired time-to-frequency conversion process.

The FWM process between a linearly chirped pump beam (with approximately constant intensity over the signal pulse duration) and the signal, generates an idler with an intensity given by that of the dispersed signal, while the chirp is compensated for by the pump pulse. Hence, when detected on a suitable optical spectrum analyzer the idler pulse, in the frequency domain, represents the temporal profile of the PUT.

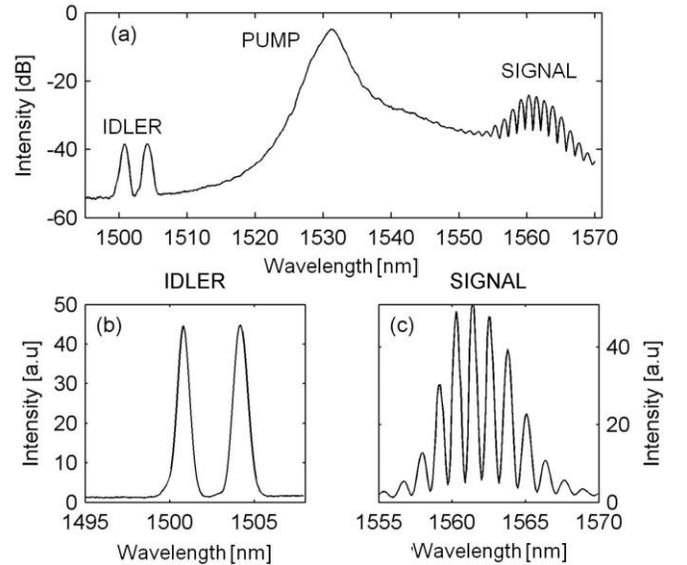

Fig. 8 (a) Optical spectrum analyzer output of a TL experiment for a pump dispersion equivalent to 200 m of SMF (dB scale). The pump was centered at 1530 nm. The spectrum of the signal (consisting of two delayed pulses) is visible at (around) 1560 nm. The interference of the two replicas is evident. The FWM idler (around 1502 nm) clearly shows an image of the PUT (double pulse). (b) Idler spectrum (linear scale). (c) Signal spectrum (linear scale).

## V. TIME TO FREQUENCY CONVERSION MEASUREMENTS: RESULTS

A sketch of the experimental setup for the time-lens measurements is shown in Fig. 7. Subpicosecond pulses for the pump and the signal were prepared from a 17MHz repetition rate mode-locked fiber source, providing pulses at λ=1550nm. The pulses were spectrally broadened after propagation in a nonlinear fiber. As in the previous FWM experiment, synchronized and coherent PUT and pump pulses were obtained by way of two tunable Gaussian bandpass filters operating in transmission with a -3dB bandwidth of 5nm (equivalent to a pulsewidth of ~700fs) and 8nm (equivalent to a pulsewidth of ~570fs), centered at 1560 and 1530nm respectively. An interferometer was used to shape the PUT in a double pulse waveform, and a movable mirror controlled their relative delay.

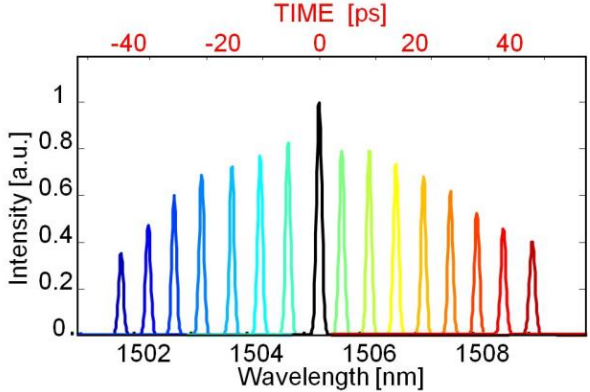

Fig. 9 Idler spectrum for a PUT consisting of two delayed Gaussian pulses, for different temporal delays. Here, the pump dispersion is equivalent to 2km of SFM. The first replica (black Gaussian pulse, at the center of the picture) does not change in time, while the second replica is recorded for different temporal delays and is represented with different colors.

The dispersion of the pulses was controlled via the use of different lengths of SMF: the signal dispersion was carefully adjusted to obtain half as much as the pump dispersion. Pulses were first amplified with a standard erbium doped fiber amplifier (EDFA), and were then coupled into the 45nm waveguide. An optical spectrum analyzer provided the output spectrum.

We characterized our system in two different regimes of pump dispersion: low dispersion (200m of SMF) and high dispersion (2km of SMF). It is important to underline that in our configuration we used SMFs as dispersive elements. For this reason we had to inject a weak pump pulse at the input of the SMF spool, to avoid self phase modulation (SPM) in the fiber. Our set-up was then limited by the maximum amount of average power (<30mW) and gain supplied by the EDFA. We used the low dispersion configuration (shorter pump pulse and higher pump peak power) to evaluate our device in terms of conversion efficiency (signal to idler), while in the high dispersion configuration we could evaluate the time-bandwidth product.

A typical result for the dispersion given by 200m of SMF fiber is shown in Fig. 8. The SPM-free pump is visible at 1530nm; the signal spectrum centered at 1560 clearly shows the interference of the two delayed Gaussian pulses used as the PUT in our experiment. In particular, the FWM product around 1502nm is the temporal image of the PUT. In this configuration the pump was stretched to 15ps, with approximately 10W peak power. We could achieve -10dB of conversion efficiency (signal to idler), and we verified the linearity of the time to frequency conversion of the system in this regime, confirming that no distortion is visible due to higher order nonlinear effects such as nonlinear absorption.

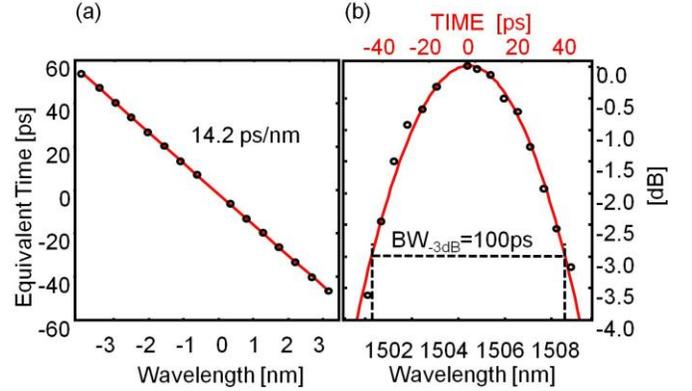

Fig. 10 Calibration curve for the delay between the two Gaussian pulses composing the PUT (shown in Fig. 9): experimental data extracted from Fig. 9 (black dots) and related linear interpolation(red line). (b) Maxima of the Gaussian pulses in Fig. 9 for different time delays (black dots) and related parabolic interpolation of the experimental maxima (red curve). We note that the intensity of the imaged pulses varies because the approximation of a constant pump intensity does not hold for the large delays used in this experiment.

Fig. 9 shows the resulting idler spectrum generated by the device as a function of the delay between the two peaks of the double pulse waveform, for a pump dispersion equivalent to 2km of SMF. In this configuration the pump was stretched to approximately 100ps and the peak power was 1W. By measuring the spectral separation of the two idler peaks (see Fig. 9), a calibration curve was obtained (see Fig. 10). The time delay between the two temporal pulses was accurately extracted from the interference fringes in the signal spectrum. The measurement shows an excellent linear trend, as expected by relation (2). We obtained a calibration factor of 14.2ps/nm. The peak value of the varying pulse replica in the measured idler spectrum vs wavelength is shown in Fig. 10 (b); with the calibration measurement in Fig. 10(a) we could calculate the corresponding temporal axis, shown at the top of Fig. 10(b), in red. This measurement allowed us to precisely estimate the temporal recording length of our instrument, obtaining a -3dB time window of 100ps.

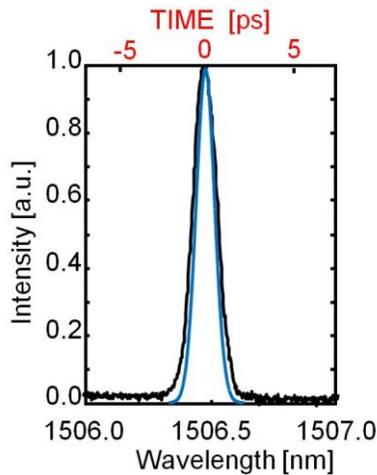

**Fig. 11 Idler spectrum (black curve, for a PUT consisting of a transform limited Gaussian pulse with 5nm bandwidth. The converted timescale (red axis at top) is obtained using the calibration factor of 14.2ps/nm extrapolated from Fig. 10. The blue curve is the FT of the signal (i.e. PUT) spectrum, representing the transform limited PUT temporal waveform.**

The calibration measurement, as visible from equation (2), provides an experimental evaluation of the actual phase curvature acting in the TL process, giving an accurate measurement of the quadratic curvature of the pump. We could then determine the correct amount of dispersion to be added to the signal to observe its temporal image in the idler spectrum. If the amount of dispersion is not accurately compensated for, then the image will appear "blurred", as occurs in the spatial equivalent case when an imaging instrument is out of focus. We controlled the signal dispersion with an accuracy of approximately 10m of SMF, due to synchronization constraints between signal and the pump. This is due to the fact that, when the length of the SMF spool used to disperse the PUT is modified, the relative delay between pump and signal is also affected.

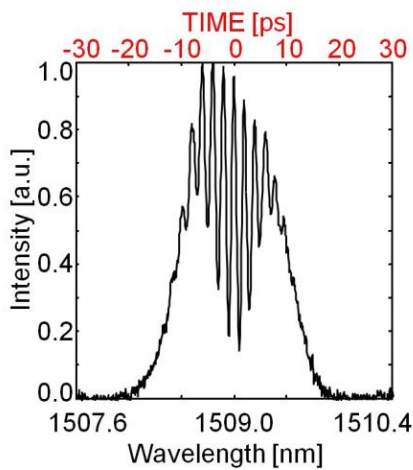

**Fig. 12 – Idler spectrum of a temporal shape with a large time bandwidth product, revealing 800fs long oscillations over a temporal duration of 40 ps.**

The image of a single, transform limited Gaussian pulse is shown in Fig. 11. It compares the result of our TL experiment with the transform limited temporal shape obtained from the FT of the spectral measurement of the signal, yielding a Gaussian pulsewidth of ~700fs. This was confirmed by autocorrelation measurements. In comparison, the TL measurement yields a pulsewidth of ~ 900fs, or roughly 25% larger. This is a result of two effects. First, we estimate that the resolution of our time lens experiment is ~ 400fs - as previously noted, a $\tau_0 = 570$fs temporal pump pulse duration in the transform limited section corresponds to a resolution of $\tau_0/\sqrt{2}$ ~ 400fs. This broadens the pulsewidth to ~ 800fs [21],[25]. The additional broadening to ~ 900fs arises from our TL system being slightly out of focus - equivalent to propagation through ~ 5m of SMF.

An instrument with a temporal resolution of 400fs over a time window >100ps possesses a time bandwidth product > 250. This is comparable with previous results obtained in silicon [18]. Better performance in terms of both output idler conversion efficiency and time-bandwidth product can be reached for larger pump excitations, limited by our set-up but not by our device. As our platform is not affected by nonlinear absorption, high power can be employed without distortion inducing aberrations in the TL. This temporal imaging instrument is also well suited for the measurement of pulses with complex temporal features over large scales. As is visible from Fig. 12, a temporal shape consisting of an oscillation with a time scale of 800fs over a temporal window larger than 40ps is successfully imaged.

## VI. Conclusion

We demonstrate all-optical temporal measurement of subpicosecond optical pulses by TL imaging, based on FWM in a 45cm long high index glass integrated spiral waveguide, achieving a time-bandwidth product better than 250 (temporal resolution ~ 400fs and measurement time window ~100ps). We also show a conversion efficiency signal to idler of -10dB in a TL configuration based on a shorter pump pulse. Our results originate from a combination of very low linear and nonlinear loss, a reasonably high nonlinearity and near ideal dispersion characteristics of the waveguide. The high stability, manufacturability and CMOS compatible fabrication of this integrated platform are attractive features for developing practical devices for systems applications, including 'on chip' ultrafast optical signal measurement platforms.